\documentclass[superscriptaddress,reprint,amsmath,amssymb,aps,floatfix]{revtex4-1}

\usepackage{amsmath,gensymb,textcomp,bm,dcolumn,eurosym,array,tabu,multirow,nicefrac,color,subfigure,graphicx,upgreek}
\usepackage[colorlinks, linkcolor=blue, citecolor=blue, urlcolor=blue, breaklinks=true]{hyperref}

 \expandafter\let\csname equation*\endcsname\relax
\expandafter\let\csname endequation*\endcsname\relax

\usepackage{inputenc}

\newcommand{\ket}[1]{\lvert #1 \rangle}

\makeatother

\begin{document}
\title{Passively stable distribution of polarisation entanglement over 192\,km of deployed optical fibre}

\author{S\"oren Wengerowsky}
  
\email{ soeren.wengerowsky@oeaw.ac.at}
\affiliation{Institute for Quantum Optics and Quantum Information - Vienna (IQOQI), Austrian Academy of Sciences, Boltzmanngasse 3, 1090 Vienna, Austria}
\affiliation{Vienna Center for Quantum Science and Technology (VCQ), Boltzmanngasse 5, 1090 Vienna, Austria}

\author{Siddarth Koduru Joshi}
\affiliation{Quantum Engineering Technology Labs, H.\ H.\ Wills Physics Laboratory \& Department of Electrical and Electronic Engineering, University of Bristol, Merchant Venturers Building, Woodland Road, Bristol BS8~1UB, United Kingdom}

\author{Fabian Steinlechner}
\affiliation{Fraunhofer Institute for Applied Optics and Precision Engineering IOF Jena, Albert-Einstein-Stra\ss{}e 7, 07745 Jena, Germany  }
\affiliation{Abbe Center of Photonics, Friedrich Schiller University Jena, Albert-Einstein-Str. 6, 07745 Jena, Germany}

\author{Julien R. Zichi}
\affiliation{Department  of  Applied  Physics,  Royal  Institute  of  Technology  (KTH),  SE-106~91  Stockholm, Sweden}
\affiliation{ Single Quantum B.V., Molengraaffsingel 10,  2629JD Delft, The Netherlands}

\author{Bo Liu}
\affiliation{Institute for Quantum Optics and Quantum Information - Vienna (IQOQI), Austrian Academy of Sciences, Boltzmanngasse 3, 1090 Vienna, Austria}
\affiliation{Vienna Center for Quantum Science and Technology (VCQ), Boltzmanngasse 5, 1090 Vienna, Austria}
\affiliation{College of Advanced Interdisciplinary Studies, NUDT, Changsha, 410073, China}

\author{Thomas Scheidl}
\affiliation{Institute for Quantum Optics and Quantum Information - Vienna (IQOQI), Austrian Academy of Sciences, Boltzmanngasse 3, 1090 Vienna, Austria}
\affiliation{Quantum Optics, Quantum Nanophysics and Quantum Information, Faculty of Physics, University of Vienna, Boltzmanngasse 5, 1090 Vienna, Austria }

\author{Sergiy M. Dobrovolskiy}
\affiliation{Single Quantum B.V., Molengraaffsingel 10,  2629JD Delft, The Netherlands }

\author{Ren\'{e} van der Molen}
\affiliation{Single Quantum B.V., Molengraaffsingel 10,  2629JD Delft, The Netherlands }

\author{ Johannes W.\ N.\ Los}
\affiliation{Single Quantum B.V., Molengraaffsingel 10,  2629JD Delft, The Netherlands }

\author{Val Zwiller}
\affiliation{Department  of  Applied  Physics,  Royal  Institute  of  Technology  (KTH),  SE-106~91  Stockholm, Sweden}
\affiliation{ Single Quantum B.V., Molengraaffsingel 10,  2629JD Delft, The Netherlands}

\author{Marijn A.\ M.\ Versteegh}
\affiliation{Department  of  Applied  Physics,  Royal  Institute  of  Technology  (KTH),  SE-106~91  Stockholm, Sweden}

\author{Alberto Mura}
\affiliation{Istituto Nazionale  di Ricerca  Metrologica (INRIM), Strada delle Cacce, 91,
10135  Turin, Italy} 

\author{Davide Calonico}
\affiliation{Istituto Nazionale  di Ricerca  Metrologica (INRIM), Strada delle Cacce, 91,
10135  Turin, Italy}

\author{Massimo Inguscio}
\affiliation{European Laboratory for Non-Linear Spectroscopy (LENS), Via Nello Carrara, 1, 50019 Sesto Fiorentino, Italy}
\affiliation{Department  of  Engineering,  Campus  Bio-Medico  University  of  Rome, Via Alvaro del Portillo, 21,  00128  Rome,  Italy}
\affiliation{Consiglio Nazionale delle Ricerche (CNR), Piazzale Aldo Moro, 7, 00185 Rome, Italy}
 
\author{Anton Zeilinger}
\affiliation{Institute for Quantum Optics and Quantum Information - Vienna (IQOQI), Austrian Academy of Sciences, Boltzmanngasse 3, 1090 Vienna, Austria}
\affiliation{Quantum Optics, Quantum Nanophysics and Quantum Information, Faculty of Physics, University of Vienna, Boltzmanngasse 5, 1090 Vienna, Austria }

\author{ Andr\'e Xuereb}
\affiliation{Department of Physics, University of Malta, Msida MSD~2080, Malta}

\author{Rupert Ursin}
\email{rupert.ursin@oeaw.ac.at}
\affiliation{Institute for Quantum Optics and Quantum Information - Vienna (IQOQI), Austrian Academy of Sciences, Boltzmanngasse 3, 1090 Vienna, Austria}
\affiliation{Vienna Center for Quantum Science and Technology (VCQ), Boltzmanngasse 5, 1090 Vienna, Austria}

\begin{abstract}
Quantum key distribution (QKD) based on entangled photon pairs holds the potential for repeater-based quantum networks connecting clients over long distance. 
We demonstrate long-distance entanglement distribution by means of polarisation-entangled photon pairs through two successive deployed $96$\,km-long telecommunications fibres in the same submarine cable. One photon of each pair was detected directly after the source, while the other travelled the fibre cable in both directions for a total distance of $192$\,km and attenuation of $48$\,dB. The observed two-photon Bell state exhibited a fidelity $85$\%$\pm 2$\,\% and was stable over several hours. We employed neither active stabilisation of the quantum state nor chromatic dispersion compensation for the fibre.
\end{abstract}

\maketitle
 
\section{Introduction}
  
The maturity of quantum communication and the information theoretic security it provides have already found multiple applications in metropolitan fibre networks~\cite{Wonfor2018} including some elections~\cite{swisselections}. 
Studying quantum communication over long-distance links marks the next step in the advancement of this technology. Notably, entanglement provides the potential for being able to generate a secure key, over longer distance than decoy based quantum cryptography~\cite{Ma2007,Scheidl2009}. 
Entanglement also facilitates device independent quantum key distribution (QKD), where a secure key can be generated even if the devices used are provided by an adversary~\cite{masanes2011_device_indep,branciard2012_semi_device_indep}. 
Further, the distribution of entanglement allows entanglement purification, which is a fundamental part of the implementation of quantum repeaters.
 Since even an ideal quantum repeater would be susceptible to loss, it is of paramount importance to demonstrate entanglement distribution over the longest distance possible.
This will enable applications such as QKD and distributed quantum computation over distances beyond the metropolitan length scale. 
Thus in the long run, we believe that entanglement distribution will play a key role in future quantum communication techniques.

Using a satellite, a distance record has been achieved both for QKD with a trusted node~\cite{liao2017satelliteqkd} as well as for the longest distance quantum entanglement has ever been deployed over~\cite{Yin2017a}, bridging a geographical distance of $1200$\,km while yielding coincident counts at a rate of $1.1$\,s$^{-1}$ over a total loss in the dual link of $64$ to $82$\,dB. If the distance were to be bridged by a fibre-optic cable connecting the two end-points directly, the attenuation would be at least $170$\,dB, assuming world-record low-loss fibres having a loss of $0.142$\,dB/km~\cite{tamura2017lowest}. Nevertheless, for links of moderate length, fibre-optic connections are often the more suitable solution, in terms of the amount of key generated per year.

The distribution of entanglement through fibre-optic cables has been predominantly focused on time-bin entanglement~\cite{Tittel1998,Marcikic2002,honjo2007_100km_timebin,Inagaki2013,Aktas2016}, where the entangled degree of freedom between the two photons is temporal, i.e., related to the order in which the photons arrive, as opposed to their polarisation. The former has been demonstrated over two 150\,km arms~\cite{Inagaki2013}, while the latter has been demonstrated over a single arm consisting of a 100\,km fibre spool~\cite{hubel2007high}. The longest distribution of polarisation entanglement over deployed fibre so far is 96\,km~\cite{Wengerowsky201818752PNAS}.

Polarisation entanglement/encoding, in particular, provides a conceptually higher key generations rate. The largest concern is its susceptibility to environmental influences like movement or temperature changes along the link. 
Fluctuations in the polarisation state transmitted through the fiber cable result in an increased quantum bit error rate (QBER) and therefore a reduction of the key rate.
Our results suggest, that polarisation encoding can be used to its full advantage, even on a deployed long-distance fibre link.

In this paper we demonstrate entanglement distribution, using the polarisation degree of freedom, through a total of $192$\,km of deployed optical fibre that is part of the active classical telecommunications network linking the Mediterranean islands of Malta and Sicily. The fibres we used are dark fibres carrying no classical signals, deployed in a bundle containing  other optical fibres carrying classical signals. Although polarisation entanglement has been demonstrated in a laboratory setting over $100$\,km of spooled fibre~\cite{hubel2007high}, and the longest deployed fibre link was $96$\,km long~\cite{Wengerowsky201818752PNAS}, it is possible to extend this range significantly. In the present work, one photon from each entangled pair was detected right at the source in Malta, whilst the second was sent through the submarine cable to Sicily and back to Malta via a second dark fibre, where it was eventually detected. Our demonstration is carried out at the same location as in~\cite{Wengerowsky201818752PNAS}. The experiment was conducted without the use of either compensation of chromatic dispersion, or any active stabilisation techniques. We conclude that polarisation entanglement is robust enough to be used in current commercial scenarios and provides an attractive alternative to time-bin entanglement over long fibre links. 

\section{Setup} 

\begin{figure*}[htp]
 \centering
     \includegraphics[width=\linewidth]{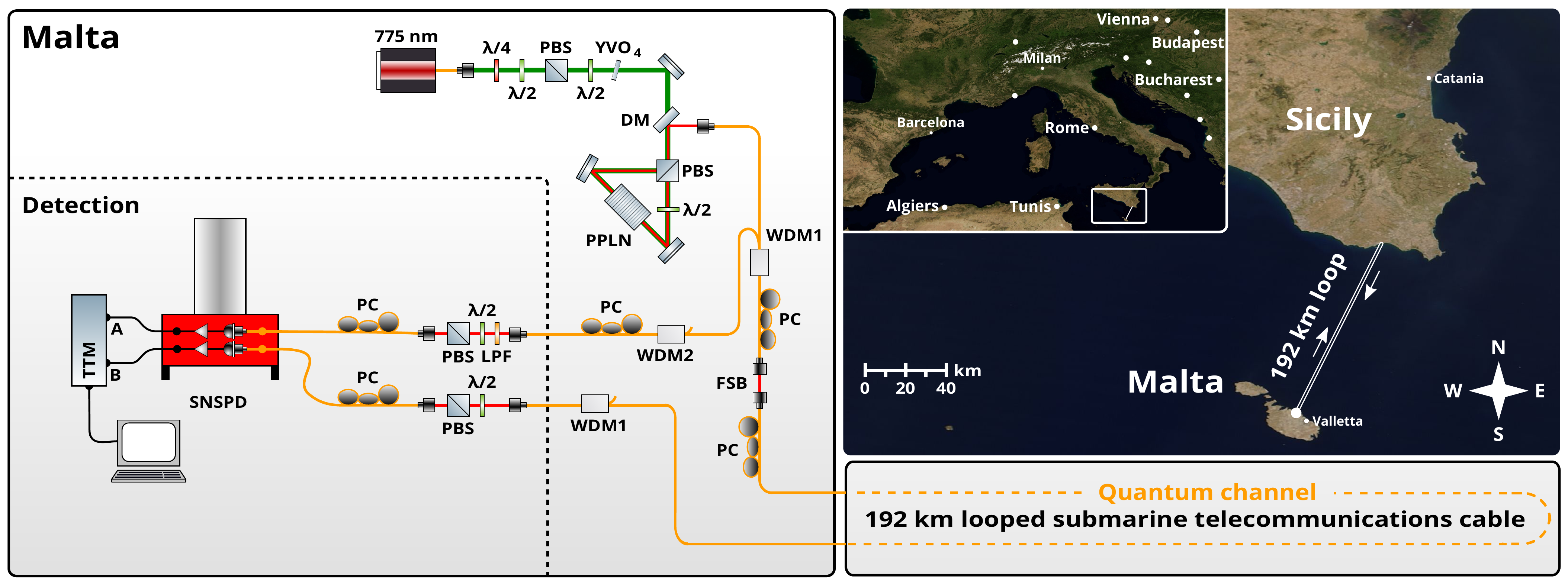}
 \caption{\label{setup_and_map}  Setup and location of the experiment: We used a fibre optic cable which links the Mediterranean islands of Malta and Sicily.  A continuous-wave laser at $775$\,nm produces, via spontaneous parametric down-conversion, photon pairs which are entangled in polarisation due to the Sagnac geometry. Signal and idler photons are separated by frequency into two different fibres; one photon is detected immediately in Malta in a polarisation analysis module consisting of a half-wave plate in front of a PBS and one superconducting nanowire single-photon detector (SNSPD), and the other detected by a second detector in the same cryostat after transmission through the 192\,km submarine fibre loop and a polarisation analysis module. (Abbreviations---$\lambda/4$, $\lambda/2$:\ wave-plates; PBS:\ polarising beam-splitter, YVO$_4$:\ yttrium orthovanadate plate; DM:\ dichroic mirror; PPLN:\ MgO-doped periodically poled lithium niobate crystal (MgO:ppLN); WDM1: 0.6\,nm band-pass filter (centre wavelength $1551.72$\,nm), WDM2:\ 0.6\,nm band-pass filter (centre wavelength $1548.51$\,nm); PC:\ fibre polarisation controllers; LPF:\ 780\,nm long-pass filter; SNSPD:\ superconducting nanowire single photon detectors; TTM:\ time-tagging module; FSB:\ free-space beam. Mirrors and fibre couplers not labelled, lenses omitted.) Photos courtesy of NASA Worldview.}
\end{figure*}

An illustration of our experimental setup is depicted in Fig.~\ref{setup_and_map}. A photon source creates polarisation-entangled photon pairs by spontaneous parametric down-conversion. Spectral filtering using wavelength-division multiplexers separates the photons into two frequency channels --  ITU Grid channels 32 ($1551.72$\,nm) and 36 ($1548.51$\,nm) -- with the spacing between adjacent channels being $100$\,GHz and with each channel having a full-width at half maximum (FWHM) of $0.6$\,nm (similar to that described in Ref.~\cite{Wengerowsky201818752PNAS}). One photon from each pair was sent to a polarisation analysis and detection module located in Malta close to the source. Its entangled partner photon was sent to Sicily via a submarine telecommunications optical fibre cable and looped back to Malta. The submarine cable, $96$\,km long each way, introduced a one-way attenuation of $24$\,dB and consisted of a bundle of several non-zero dispersion shifted fibres (Corning LEAF~\cite{CorningLEAF}) which comply with the specifications of the International Telecommunication Union (ITU-T G.655). The cable contained some fibres carrying classical data in the C-band around $1550$\,nm at optical powers in the order of milliwatts, as well as two dark fibres that were used as the quantum channel. In Sicily the two fibre ends were patched together in an underground utility vault on the outskirts of the town of Pozzallo (Italy), resulting in a looped quantum channel $192$\,km in length, over which the total attenuation was measured at $48$\,dB. Although the fibres were located inside the same cable, no cross-talk from the classical signals was observed in the quantum channel at our wavelengths. Upon its return to Malta, the photon was detected in close proximity to the source.

\subsection{The polarisation-entangled photon-pair source}
Our source was based on type-$0$ spontaneous parametric down-conversion in a $4$\,cm-long magnesium oxide doped periodically poled temperature stabilised lithium niobate (MgO:ppLN) bulk crystal with a poling period of $19.2$\,$\upmu$m. The type-$0$ process converts, with low probability, one linearly-polarised pump photon at $775.075$\,nm from a CW laser to two daughter photons, commonly referred to as the signal (s) and idler (i) photons, in the C-band, having the same polarisation as the pump photon. The MgO:ppLN crystal was bidirectionally pumped inside a Sagnac-type setup~\cite{Kim2005} including a half-wave plate, thus creating a polarisation-entangled state.  

\begin{equation}\label{eq:state}
\ket{\Phi^{-}}=\tfrac{1}{\sqrt{2}}(\ket{\text{V}_\text{s}\text{V}_\text{i}} - \ket{\text{H}_\text{s}\text{H}_\text{i}}),
\end{equation} 

The birefringence of the quantum channel was  compensated for in both the H--V (horizontal--vertical) and the D--A (diagonal--anti-diagonal) bases by sending laser light at $1551.72$\,nm through the quantum channel from the point in the setup corresponding to where the SNSPD system lies in the figure, and analysing it with a polarimeter in the region labled "FSB" in fig~\ref{setup_and_map}. With the manual polarisation controllers, the fibre was aligned such that the polarisation state of the light, as defined by the polarising beam splitter and half-wave plate in front of the detector also arrives at FSB. Following this, the laser was disconnected again and the SNSPD was connected in its stead, allowing quantum measurements to be performed in the configuration depicted in Fig.~\ref{setup_and_map}. 

\subsection{The detection system}
The photons were detected using two separate fibre-coupled superconducting nanowire single-photon detectors (SNSPD) in the same commercial croystat (Single Quantum Eos) operating at $2.9$\,K. A current driver (Single Quantum Atlas) was used to read out the signals, which were digitised by a time tagging device.  Since the efficiency of the detectors is dependent on the photon polarisation, we used a manual fibre polarisation controller to optimise the detection efficiency. Detector ``A'' exhibited a dark-count rate of $900$ per second at an efficiency of $60$\%, whereas detector ``B'' was operated at an efficiency of $12$\% to reduce the dark-count rate to $20$ clicks per second.

\begin{figure}[t]
 \centering
 \includegraphics[width=\columnwidth]{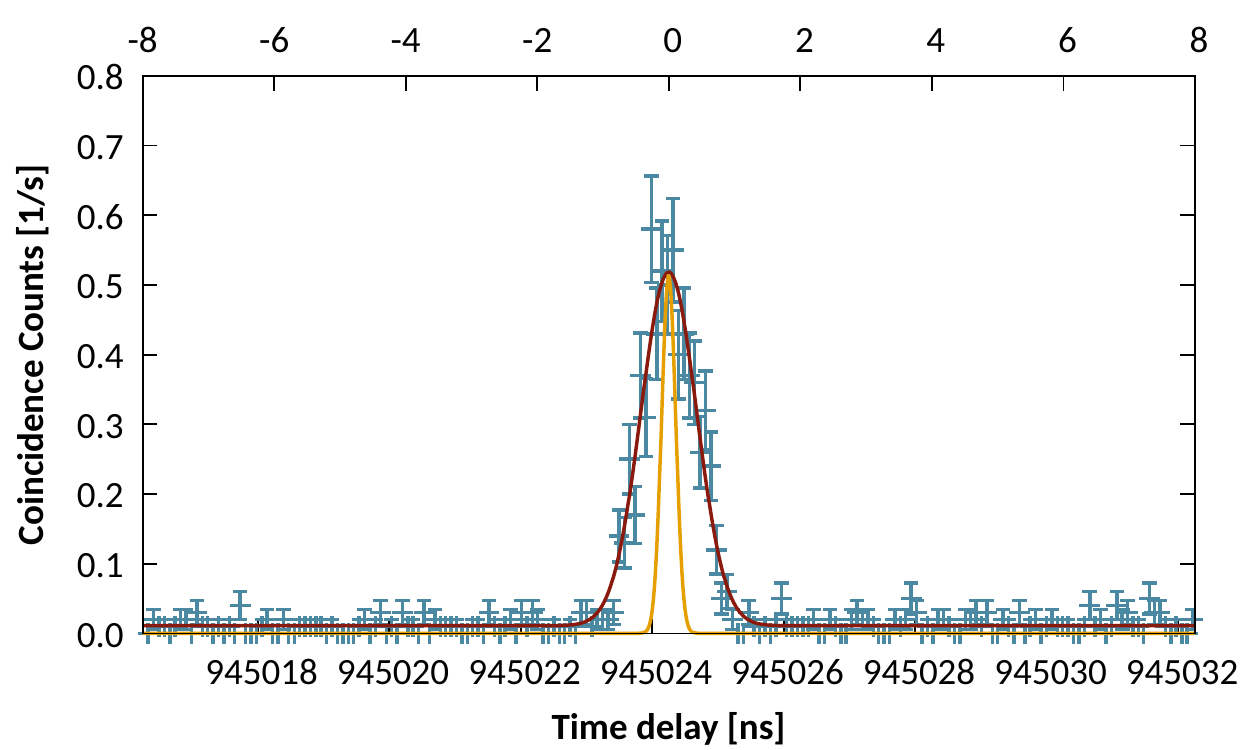}
 \caption{\label{fig_g2peak192} The cross-correlation function between the time tags from the two detectors shows a peak at a relative delay of $0.945$\,ms, which corresponds exactly to the total loop distance of (192820.538 $\pm$ 0.004)\,m, assuming that the latency of the two detectors is identical. The full-width at half-maximum is $1$\,ns. The three main factors that contribute to the peak width are the chromatic dispersion of the fibre link of $760$\,ps for our signal spectrum, the timing jitter of the detectors and of the time-tagging units -- $\sim250$\,ps and the clock synchronisation accuracy estimated from a separate measurement (using a 1\,ms optical delay in the lab) of about 500\,ps.  The data shown here was measured over $100$\,s, while the rate of coincident clicks was $3.8\pm0.2$\,s$^{-1}$. The orange graph is a fit to a measurement using the same detectors without the long fibre link. It has been normalised to the same height as the measured counts from the link experiment. The FWHM is here $250$\,ps in this case.}
 
\end{figure}

\begin{figure*}[t]
 \centering
 \includegraphics[width=1.7\columnwidth]{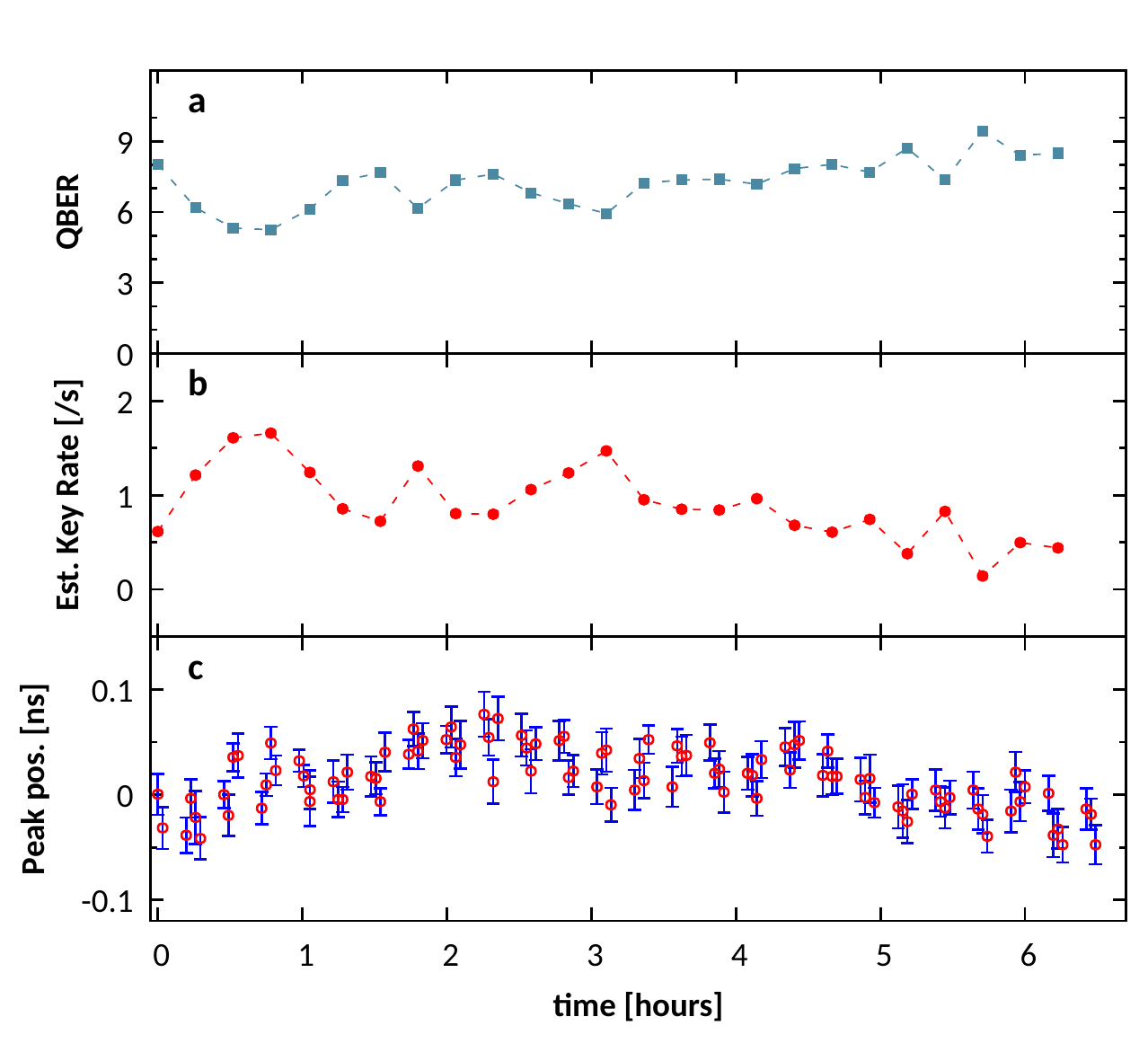}
 \caption{\label{fig_vis192} Long-term measurements on the entangled photon state over the $192$\,km fibre link, demonstrating excellent passive stability of the system. \textbf{a}~QBER, estimated from the visibility measurements. The QBER shown were calculated from 8 different combinations of measurement bases each integrated over 100\,s (The third and fourth data points correspond to a measurement duration of $84$\,s and $80$\,s long, respectively).
   \textbf{b}~The red points correspond to the estimated secure key rate, based on the observed count rate and fidelity in the asymptotic time limit.
 \textbf{c}~Position of the coincidence peak, as observed from histograms similar to Fig.~\ref{fig_g2peak192} over an interval of several hours during which no compensation of the fibre birefringence was carried out. The error bars correspond to one standard deviation as derived from the fitting parameters. }
\end{figure*}

\section{Results and discussion}
Pairs of photons were identified using timing cross-correlations, an example of which is shown in Fig.~\ref{fig_g2peak192}. The large effective jitter of the system was a primary source of error. In  an independent measurement without a long fibre, this peak is $250$\,ps wide, mainly due to timing uncertainties in the time-tagging unit and the jitter of the detection system. Improved electronics could more than halve this. Chromatic dispersion accounted for a further $760$\,ps~\cite{CorningLEAF}, taking into account the shape of the spectrum. This effect can also be mitigated, for example by using narrow band chirped fibre Bragg grating filters~\cite{Hill:94}. Further, separate laboratory experiments have shown that the synchronisation scheme employed using the time taggers introduced a  further $500$\,ps of timing uncertainty when measuring a temporal delay of about $1$\,ms. This effect could also be compensated by locking the internal clock of the time tagging unit to a more stable external clock.  
The above techniques should enable us to significantly increase the link distance and key rate.

The number of coincident pairs of photons was calculated by integrating the correlation function over a region of $823$\,ps, equivalent to $10$ time-bins of the time tagging unit. The local count rates were $2.1\times10^6$\,s$^{-1}$ in the first detector and $55$\,s$^{-1}$\,$\pm$\,$2$\,s$^{-1}$ in the detector after the link, with $20$\,s$^{-1}$ of these being dark counts. The average rate of coincidence counts for the $100$\,s-long measurements was $4.3\pm 0.3$\,s$^{-1}$ for measurements that were supposed to be correlated (H--V, V--H, D--A, A--D) and $0.3\pm0.2$\,s$^{-1}$ for measurements where no correlation was expected.

To quantify the quality of the entangled state after transmission through both fibres used in the submarine cable, we performed a series of two-photon correlation measurements. The polarisation-analysis modules were used to measure the coincidence visibility in the H--V or D--A bases, with four measurements required for each basis setting. The fidelity of the two-photon state with respect to Bell state (Eq.~\ref{eq:state}) is lower-bounded by the arithmetic mean of the two visibilities~\cite{Blinov2004,Chang2016}. The highest fidelity measured 
was $88\%\pm2\%$. Locally, the fidelity was characterised to be $98\%\pm0.2$\%. This apparent deterioration of the quantum state is attributed mainly to the rather large coincidence window of $823$\,ps (which was a compromise between the optimum pair rate and least error rate) as well as the high local count rate of $2.1\times10^6$\,s$^{-1}$.   This, together with the dark counts, increases the accidental coincidence count rate and deteriorates the detected quantum state. At the given count rate and local visibility, the best fidelity that can be achieved with this current system and coincidence window is approximately $93.5$\%, since accidental coincidence clicks deteriorate the quality of the measured quantum state. The discrepancy of roughly $5$\% between this value and our measured results is attributed to the imperfect fibre birefringence compensation and other systematic effects such as offsets in the calibration of the polarisation analysis modules. Previous studies concerning polarisation-mode-dispersion suggest that a deterioration in fidelity by only a few tenths of a percent~\cite{antonelli2011sudden,brodsky2011loss} is expected for situations such as ours, where about $0.6$\,ps of polarisation-mode-dispersion is expected~\cite{CorningLEAF}; however, this is clearly not the dominant effect in our case.
 
In order to measure the stability of our setup, including the submarine fibre link, we performed a long-duration measurement of the visibility over a period of about $6.5$\,hours. Throughout this period, the visibilities in the H--V and D--A bases were measured alternately. Imperfect compensation of the quantum channel and other systematic issues mean that the visibilities of the two bases were quite different. In the H--V basis, the visibility ranged from $74\pm2$\% to $86\pm2$\%, which corresponds to a maximum polarisation rotation on the Poincare sphere of $12{\degree}$. The visibility in the D--A basis was slightly more stable, as it ranged from $87\pm2$\% to $94\pm2$\%, corresponding to a maximum polarisation rotation of $9\degree$. Nevertheless, our results show that the lower bound to the fidelity hardly changes over the course of the six hours. The measured average fidelity was $85$$\pm2$\%, well above $1/\sqrt{2}\approx70.7$\%, which is the minimum fidelity required to violate a CHSH inequality and therefore certify entanglement, for the duration of the experiment. For entanglement-based QKD protocols, a fidelity of $81$\% is required to yield a positive secret key rate~\cite{Ma2007a}. This more stringent bound is also surpassed in our data.
Since only two detectors were available to us, without a fast basis choice, in total eight measurements were needed to estimate the fidelity. We use these results to estimate the sifted key rate. A setup which chooses a random basis with 50\% probability would measure all these measurement combinations in the same time interval, together with another eight combinations, in which both users measure in a different basis. Therefore, the fast switching QKD setup would be able to analyse sixteen combinations of polarisation during the time in which a two-detector setup, like ours, can measure only one combination. However, the rate of coincident photons in each basis combination would only be one quarter of the rate of the two-detector setup, since the setups will only measure both in the D--A and both in the H--V for one quarter of the time, respectively. Therefore, the sifted key $R_s$ which would have been observed in a fast-switching QKD setup is estimated to be one fourth of the sum of coincident clicks of our eight measurements. In Fig.~\ref{fig_vis192}\textbf{b}, the red points correspond to an estimate of the secure key rate, based on the observed count rate and fidelity~\cite{Ma2007} in the asymptotic temporal limit, therefore ignoring finite size effects. For this, an error correction efficiency of 1.15 was assumed~\cite{elkouss2009efficient}.

Another observable that we have access to is the relative delay between the two photons, corresponding to the length of the optical link. We used this data to assess the stability of the net optical length of the fibre by tracking the evolution of the temporal position of the coincidence peaks, which were determined via Gaussian fits similar to the one shown in Fig.~\ref{fig_g2peak192}. The error bars have been obtained from the covariance matrix of the corresponding Gaussian fits. The maximum change of the coincidence delay, $124$\,ps, between hours $2.28$ and $6.28$ can be explained by a net change in the average temperature of the fibre, which we consider a proxy for the temperature of the seabed. We estimated the temperature change to be about $22$\,mK using a thermal expansion coefficient of optical fibres of $\frac{\mathrm{d}L}{\mathrm{d}T}=5.6\times10^{-7}\,\text{K}^{-1}$ and the change of refractive index per kelvin of $\frac{\mathrm{d}n}{\mathrm{d}T}=8.45\times10^{-6}\,{\text{K}^{-1}}$~\cite{leviton2006temperature}. This agrees roughly with the findings of Ref.~\cite{clivati2018optical} where the environmentally induced phase noise in submarine fibre cables was investigated in this region. This temperature change corresponds to a length change of approximately $2.4$\,mm, which does not have any significant effect on the transmitted fidelity since slow length changes do not exert sufficient strain to cause significant birefringence. When compared to Refs.~\cite{waddy2005polarization,ding2017polarization}, our results show that this submarine environment is very favourable for polarisation-based quantum communication in optical fibres, demonstrating even greater stability than in laboratories with conventional climate control, which often only regulates the temperature with the accuracy of $1$ or $2$ kelvin.

\section{Conclusions}
We have demonstrated the distribution of polarisation-entanglement over a total length of $192$\,km using a dark fibre that is deployed in a submarine cable in close proximity with fibres actively carrying classical communication signals. Our demonstration does not involve any active stabilisation, in spite of which we observe remarkable passive stability for the polarisation state over several hours. Our work extends the length over which polarisation entanglement distribution is proven to be possible using entangled photon pairs and heralds the use of polarisation entangled photons and submarine telecommunication optical fibre links as building blocks for a quantum internet.

\noindent\textbf{Data availability.}
The data that support the findings of this study are available from the corresponding authors on request.

\noindent\textbf{Competing interests.}
The authors declare no competing financial interests.

\noindent\textbf{Author Contributions.}
The experiment was conducted by SW and SKJ. The setup was designed by FS, SW, and SKJ. JRZ, SMD, RvdM, JWNL, VZ and MAMV  helped with the detection of the single photons. AM, DC and MI helped to conduct the experiment. AX analysed the data, helped to conduct the experiment, and oversaw the work in Malta. LB and TS helped to analyse the data. AZ and RU contributed to the experimental design and source as well as supervising the project.  The paper was written by SW, FS, SKJ, and AX. All authors discussed the results and contributed to writing the manuscript.

\section*{Acknowledgements}
We are deeply indebted to Simon Montanaro, Roderick Cassar, and Charles Peresso at Melita Ltd.\ for providing assistance and access to their network.  We thank Jesse Slim for programming a user interface for our motorised rotation stages, Lukas Bulla, Matthias Fink,   Thomas Scheidl, Aron Szabo, Leah Paula Vella, and Ryan Vella for helpful discussions and technical assistance.  
We gratefully acknowledge financial support from the Austrian Research Promotion Agency (FFG) Agentur f\"ur Luft- und Raumfahrt (FFG-ALR contract 844360 and FFG/ASAP:6238191 / 854022), the European Space Agency (ESA contract 4000112591/14/NL/US), the Austrian Science Fund (FWF) through (P24621-N27) and the START project (Y879-N27), as well as from the Austrian Academy of Sciences, the European Research Council under the grant agreement No.\ 307687 (NaQuOp),  the Swedish Research Council (grants 638-2013-7152 and 2016-04527), the Linneaus Center in Advanced Optics and Photonics (Adopt), the University of Malta Research, Innovation \& Development Trust (RIDT).  We acknowledge the use of imagery from the NASA Worldview application (\url{https://worldview.earthdata.nasa.gov/}), part of the NASA Earth Observing System Data and Information System (EOSDIS)

\end{document}